# Quantum effects in the radial thermal expansion of bundles of single-walled carbon nanotubes doped with $^4$He.

A.V. Dolbin[1], V.B. Esel'son[1], V.G. Gavrilko[1], V.G. Manzhelii[1], N.A. Vinnikov[1], S.N. Popov[1], B. Sundqvist[2].

[1] B. Verkin Institute for Low Temperature Physics & Engineering NASU, Kharkov 61103, Ukraine

[2] Department of Physics, Umea University, SE - 901 87 Umea, Sweden

Electronic address: dolbin@ilt.kharkov.ua

**Abstract**

The radial thermal expansion $\alpha_r$ of bundles of single-walled carbon nanotubes saturated with $^4$He impurities to the molar concentration 9.4% has been investigated in the interval 2.5-9.5 K using the dilatometric method. In the interval 2.1-3.7 K $\alpha_r$ is negative and is several times higher than the negative $\alpha_r$ for pure nanotube bundles. This most likely points to $^4$He atom tunneling between different positions in the nanotube bundle system. The excess expansion was reduced with decreasing $^4$He concentration.

**Introduction**.

Carbon nanotubes, noted for their unique geometry and unusual physical properties, are promising materials for technological applications and fundamental research on low-dimensional objects. Previously [1-4] we used the dilatometric method to investigate the low temperature radial thermal expansion of carbon nanotube (CNT) bundles and the effect of atomic (Xe) and molecular ($H_2$, $N_2$) impurities on it. A number of effects have been detected, which are induced, on the one hand, by low frequency vibrations of individual CNTs and, on the other hand, by the collective behavior of the tubes forming a bundle. The vibrational spectrum of CNT bundles is also affected by impurities localized both at the surface of bundles and inside them. Since CNT walls form a quasi-two-dimensional system, their transverse vibrations perpendicular to the nanotube surface are characterized by negative Gruneisen coefficients [5] and hence make a negative contribution to the radial thermal expansion of the nanotubes. This contribution is dominant at the lowest temperatures [1,2]. In pure CNT bundles it leads to negative values of the radial thermal expansion at T = 2-5 K. It has been found [2-4] that saturation of CNT bundles with gas impurities suppresses the negative contribution and increases drastically the radial thermal expansion coefficient $\alpha_r(T)$ at $T \geq 3$ K. It was assumed that the impurity molecules deposited on the CNT surface and inside the tubes forming a bundle could change the structure of the system and hence the behavior of its properties from two-dimensional to three-dimensional. The impurity molecules dampen the low frequency transverse vibrations of the CNT walls and thus suppress the negative contribution to the thermal expansion. It was found [2-4] that the temperature dependences $\alpha_r(T)$ taken on CNT bundles saturated with gas impurities had peaks which were attributed to a spatial redistribution of the impurity molecules in the grooves and on the surface of CNT bundles.

It is known that the $^4$He impurity has quite a significant effect on the thermal and structural properties of carbon nanomaterials, including fullerite $C_{60}$ [6].

In this study we investigated the radial thermal expansion of bundles of single-walled carbon nanotubes with closed ends (c-SWNTs) saturated with 4He using the dilatometric method. The temperature interval was 2.1-9.5 K.

**Experimental technique.**

The radial thermal expansion of CNTs with closed ends saturated with $^4$He was investigated on the sample that was previously used to obtain the radial thermal expansion coefficient of pure SWNTs and gas-saturated CNTs [1-4]. The measurements were performed using a low temperature capacitance dilatometer (its design and measurement technique are described in [7]). The technique of preparing a sample of compacted CNTs for dilatometric investigations is described in [1]. The sample was obtained by compacting (P= 1,1 GPa) a CNT powder (Cheap tubes, USA, CCVD Method). Applying pressure to a thin (<0.4 mm) CNT layer aligns the CNT axes in the plane perpendicular to the pressure direction, the average deviation of the CNT axes from the plane ~4º [8]. The sample produced by successive compression of thin layers was a cylinder 7,2 mm high and 10 mm in diameter.

Immediately prior to the investigation, the cell with a sample of pure CNTs was evacuated at room temperature for 72 hours to remove possible gas impurities. Then $^4$He gas was fed to the measuring cell up to P=28 Torr. The filled cell was cooled slowly (for 12 hours) to T=2 K. On cooling in the course of $^4$He sorption by the CNTs, additional portions of $^4$He were added to the cell to maintain the pressure in the cell no higher than the saturated vapor pressure of $^4$He at this temperature. This saturation procedure permitted, on the one hand, filling all possible positions accessible to sorption in CNT bundles, and, on the other hand, avoiding $^4$He vapor condensation on the cell walls. Saturation was stopped after the equilibrium pressure 0.01 Torr was reached in the cell. Then the cell was cooled to T=2 K.

The further measurement of the thermal expansion was made with a vacuum of up to $10^{-4}$ Torr in the cell. Desorption investigations by the procedure described in [2] have permitted estimation of the $^4$He concentration in the CNT sample, which was 9.4 mol % under these particular conditions of saturation (mol.% is the percent relation between the number of $^4$He impurity atoms and the number of carbon atoms forming CNTs.)

**Results and discussion.**

The temperature dependence of the radial thermal expansion coefficient of the $^4$He-SWNT system taken in the interval 2.1-9.5K is shown in Fig.

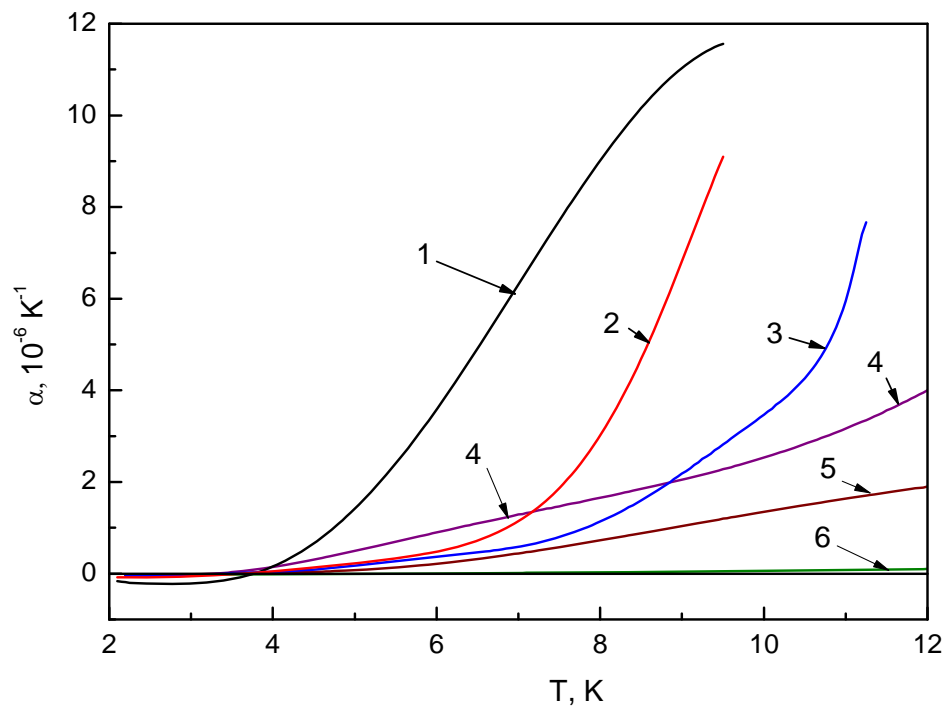

a)

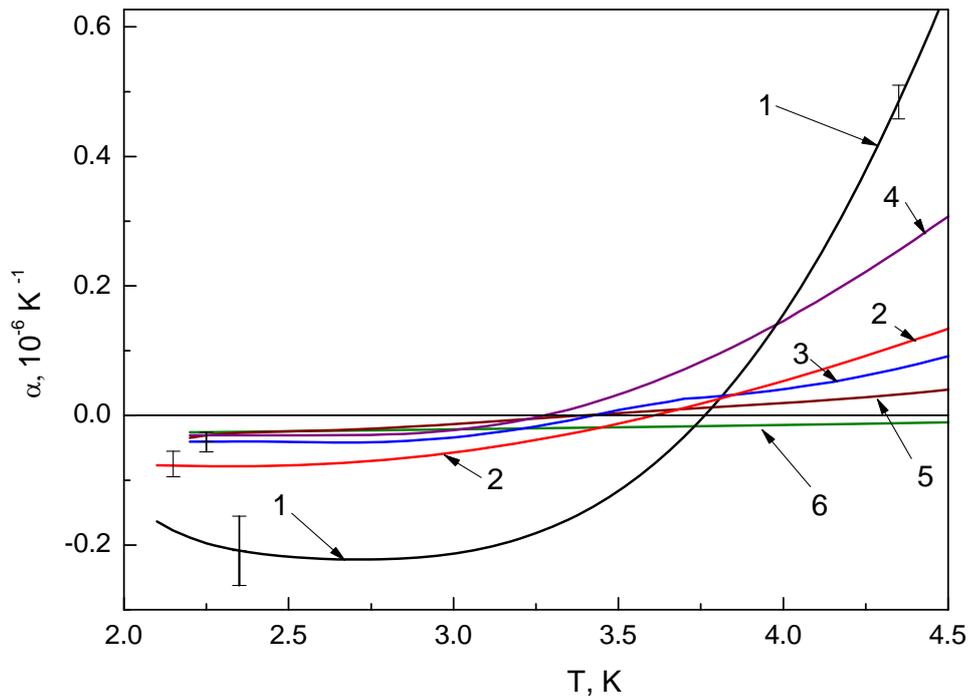

b)

Fig. Radial thermal expansion coefficient of $^4$He-saturated CNT bundles:
1 – $^4$He-SWNT, molar $^4$He concentration is 9.4%;
2 – after a partial removal of the $^4$He impurity at T=9.5K;
3 – after repeated partial removal of the $^4$He impurity at T=12K.
4 – $H_2$- SWNT [3];
5 – Xe-SWNT [2];
6 – pure SWNT [1]; at T=2.1 – 12 K (a) and 2.1 – 4.5K (b)

Since the thermal expansion was measured in the vacuum ~ $10^{-4}$ Torr, it was important to control the possible influence of $^4$He desorption on the values of thermal expansion in the course of heating the $^4$He-SWNT system above 2K. The reproducibility of the results measured on heating and cooling was continuously checked for ΔT (ΔT=0.3,…,1K). If the results obtained on the temperature cycling coincide within the experimental error, the $^4$He desorption effect can be disregarded and the results obtained more taken as equilibrium. Disturbance of the reproducibility on cycling suggests that at this and higher temperatures the $^4$He desorption from the sample affects the thermal expansion. Reproducibility of the LTEC values was observed in the interval 2.1 – 9 K and was disturbed only after heating the sample to 9.5 K. The measurements of the thermal expansion were ceased when the reproducibility was disturbed. The sample was kept at 9.5K for an hour. After a partial desorption of $^4$He the sample was re-cooled to the lowest temperature (2.1K) and the radial thermal expansion was measured again. The obtained series is described by curve 2 in Fig. To repeat a partial desorption, the sample was kept at 12 K for an hour. The dependence $α_r$(T) obtained after the repeated $^4$He desorption is described by curve 3 in Fig.

It is interesting that when the sample had the starting molar $^4$He concentration 9.4% the obtained data for $α_r$(T) exhibited very high negative $α_r$-values in the interval 2.1 – 3.7K. These values are several times higher than the corresponding negative values for pure CNT bundles (cf. curves 1 and 6 in Fig. b). As $^4$He desorption increases, the magnitude of the negative $α_r$(T) decreases along with the temperature region where the negative contribution to $α_r$(T) was dominant. Note that on measuring $\alpha_r$ of the $^4$He-SWNT solution with the concentration 9.4 mol% $^4$He the data scatter and hence the measurement error were much higher (up to five times) than the corresponding values for CNT bundles, both pure and saturated with other gases (see Fig.b). As the concentration of $^4$He in $^4$He-SWNT decreases the measurement error also decreases, approaching that for pure CNT bundles.

The results obtained in our previous studies [2-4] show that the sorbed gas should decrease the negative $\alpha_r$, or at least keep it from increasing. On the other hand, the experimental results on the thermal expansion of the $^4$He-SWNT system are taken in equilibrium, and the high negative values for $\alpha_r(T)$ cannot be explained adequately by $^4$He desorption from the sample in the course of measurement. It is natural to attribute the observed low temperature effect to tunneling of the $^4$He atoms between different positions in the CNT bundle system. According to [9], such tunneling leads to negative thermal expansion of the $^4$He-SWNT system. Since the probability of tunneling is much lower for molecules and atoms of other gases, this effect was absent in the thermal expansion of solutions of $H_2$, $N_2$ and Xe in CNT bundles [2-4].

It should be noted that the temperature dependence of the thermal expansion coefficient of SWNT bundles saturated with $^4$He contains no peaks such as were observed on saturating bundles with Xe, $H_2$ and $N_2$ [2-4]. These peaks were attributed to the thermally activated spatial rearrangement of the gas molecules (atoms). If the spatial rearrangement of the $^4$He atoms adsorbed by CNT bundles proceeds through tunneling, no peaks appear in the temperature dependence $\alpha_r(T)$.

The authors are indebted to Prof. A.S. Bakai, Prof. L.A. Pastur and Prof. M.A. Strzhemechny for fruitful discussions and to the Science & Technology Center of Ukraine (STCU) for the financial support of the study (Project # 4266).